\documentclass[runningheads,a4paper]{llncs}
\usepackage{url}
\usepackage{mathtools}
\usepackage{qtree}
\usepackage{color}
\usepackage{calc}
\usepackage{amssymb}
\usepackage{amstext}

\begin{document}


\title{Should I quit using my resource? \\ Modeling Resource Usage through Game Theory}

\authorrunning{Paraskevas V. Lekeas}
\titlerunning{Modeling Resource Usage through Game Theory}

\author{Paraskevas V. Lekeas}
\date{}

\institute{Department of Applied Mathematics,\\
University of Crete, Crete, Greece\\
\email{plekeas@gmail.com}
\\
}

\maketitle


\begin{abstract}
Existing web infrastructures have supported the publication of a tremendous amount of resources, and over the past few years Data Resource Usage has become an everyday task for millions of users all over the world. In this work we model Resource Usage as a Cooperative Cournot Game in which a resource user and the various resource services are engaged. We give quantified answers as to when it is of interest for the user to stop using part of a resource and to switch to a different one. Moreover, we do the same from the perspective of a resource's provider.


\end{abstract}

\section{Introduction}

Data Resource\footnote{When we refer to a resource we have in mind that in the background there exists a set of electronic mechanisms or internet infrastructures that created this resource for the purpose of value generation (either a profit when there is an underlying company or some other social gain, such as \cite{NamUS}). See also \cite{AtaxonomyForResourceDiscovery} for a related taxonomy.} Usage is an everyday task for millions of users all over the world. Exchanging information, communicating, working and various other aspects of our life have been inevitably affected by data repositories which can be accessed through various channels, such as the Web and the Internet via different technologies, interfaces and infrastructures \cite{TheWorld'sTechnologicalCapacitytoStoreCommunicateandComputeInformation}. Usually once someone has identified an appropriate resource, he interacts with it by exchanging information. This sort of interaction is heavily commercialized, and a huge industry\footnote{End-user spending for IT services worldwide estimated to be \$763 billion in 2009 \cite{MarketShareAnalysis:ITServicesRankingsWorldwide2009}.} has been established, which invests a great amount of money in marketing web services and products that provide access to resources quite often freely. This is why from now on we will use the word "provider" to refer to the underlying structure responsible for a resource. These providers most of the time are extremely interested in developing integrated resource services\footnote{We prefer this term instead of the term "web service" since many other alternative channels exist like satellite and cellphone grids, ad hoc networks, etc.} in order to attract users, and, more importantly, to convince them to keep using these. This is because users are valuable: They provide information to the resource by interacting with it, they bring money through the adds or subscription fees, and of course they bring new users to broaden the profit cycle.

A living example is Google which proposes a web resource experience through the integration of different technologies in order for users to continue using its services. Opening a Google account is a fairly easy one-minute process and instantaneously the new user has access to different cutting-edge services like Android OS apps, adaptive web search through the Google search engine, cloud services, access to landline phone calls, teleconferencing services and much more. A "perfect" user of Google would be the one who uses all these services explicitly through the Google APIs, sharing no data with any other competitive resource (e.g. AWS \cite{AmazonWebService} or Ubuntu One \cite{UbuntuOne}) and thus enriching only Google's resource knowledge repositories. However, many times it is the case that not all services or technologies of a resource are welcomed by users and sometimes users tend to accept only specific services from a resource ignoring some others. Also a situation that is not so good for a provider is the case where users decide to quit its resource and switch to a different one that provides similar or better services \cite{HotTodayGoneTomorrowOnTheMigrationOfMySpaceUsers}.

In this work we investigate the following problem. When do users tend to partially\footnote{Partially means that the user is unsatisfied only with some of the services and wants to switch but likes the rest and wants to keep them.} abandon a specific resource? What can resource providers do about that? Is there a way to formulate the above trends in order to be evaluated and measured? In order to approach the above questions we model the various user - service interactions within a resource with different plays of the user, which are engaged either in a cooperative or in a non-cooperative manner. Each of these plays generates a value, which is to be conceived as a measure of the user's satisfaction for the appropriate service. 

In the rest of the paper we proceed as follows: Section 2 gives a motivating example and formulates Resource Usage as a cooperative Cournot game. Section 3 studies the cases of partial rejection of a resource and also the possible reactions of the provider to prevent that. Section 4 concludes with a discussion and future work.

\section{Modeling Resource Usage}

Before describing our model let us give a stimulating example.

\subsection{The case of Zoogle+ resource}

Imagine the following scenario\footnote{Any explicit or implicit references to facts or persons is accidental and imaginary. Beware also that in Greek, Zoogle is pronounced almost the same as the Greek word $"Zo\grave{\upsilon}\gamma\kappa\lambda\alpha"$ which means Jungle referring to the chaotic and controversial informational nature of the WWW.}: Zoogle Inc. decides to offer a new web integrated data resource, named Zoogle+ that will provide its users with the following set of services: $N=\{\textit{email}, \textit{cloud}, \textit{voip}\}$. A user $u$ decides to try Zoogle+ for a certain period of time, and for this reason he signs up creating an account. Since $u$ wants to be accurate in his calculations he uses a worth function $v(\cdot)$ to rate how good his experience is. Since it is Zoogle+'s policy to prohibit the exclusive use of only one service, ignoring the rest, the single use of a service for $u$ is worth 0, i.e. $v(\{i\})=0$, $\forall \, i \in N$. Moreover, when $u$ uses any two of the services he rates his experience with a value of 2, i.e. $v(\{i,j\})=2$, $\forall \, i,j \in N$ with $i \neq j$. Finally, when using all three of the services he calculates a value of 2.5, i.e. $v(N)=2.5$. What should the user do? Should he be totally loyal to Zoogle+ or not? 

On the one hand, when $u$ uses all the services he gets on average 0.83 for each one, but, on the other hand, when he selects only two of the services he gets 0 for the one not selected and 1 for each of the rest. Therefore, $u$ decides to maximize his satisfaction and thus selects to use only two services of Zoogle+ and to seek the third in an external resource. So the answer in this case is that $u$ will partially leave. What would happen if $v(N)$ is worth 3 to $u$? Clearly then he should stay loyal to Zoogle+ because he would maximize his satisfaction in this case, since there is no combination of services that would give him more than 1 (on average).  

Underlying the intuition of the above example is the idea of the core in cooperative game theory \cite{ACourseInGameTheory}. Briefly speaking, a cooperative game is a situation in which a group of $N$ players, by making decisions that take into account each other's actions and responses, decides to act together to generate some value. This value (or what is left if we put aside the costs of playing the game) must later be split according to certain rules agreed upon by these players. The various ways this split can be made are defined as solution concepts. One of the most used solution concepts is that of the core. According to this, the split is in the core if no subset of players can benefit (earn more) by breaking away from the whole set $N$. Under this light, user $u$ decided to partially leave Zoogle+ when $v(N)=2.5$ because the core of the game was empty, while when $v(N)=3$ the core became non-empty and thus $u$ used all the services. In what follows we formulate Resource Usage as a cooperative Cournot game.

\subsection{The model}

Suppose that a user $u$ decides to sign up for a resource $R$ in order to \textit{use} its $i$, $i \in \{1,\cdots,n\}$, different services. The phrase "use a service" refers to the interaction of $u$ with the service in order for some desired tasks to be completed. For example, using the GUI of a service, entering data by typing, downloading files, writing scripts and compiling them online can be perceived as parts of such an interaction. Let $p_i$ denote the interaction of $u$ with service $i$. Call each $p_i$ a \textit{play} that $u$ does with the service. Every play\footnote{from now on the terms "play" and "interaction" will be used interchangeably.} generates some value for $u$. Since $u$ needs to make an effort to generate this value we assume that the \textit{per unit cost} to $u$ for playing $p_i$ is $c$. For example, if a play generates 2 units of value for $u$, then the cost of the play would be $2c$. In general the cost function for $q_i$ units of value produced in a play would be $cq_i$. We assume for clarity of the exposition that $c$ is the same for every $p_i$, $i=1,\cdots,n$. If we now take the value generated from a play and subtract the cost spent to produce it, then we will find how much the play is \textit{worth} to the user, or in other words we will find the profit of the user from using the specific service, which intuitively represents a measure of how satisfied the user is with the service. So each play contributes some worth to the user and if we add all these contributions from all the plays we will have the total worth to the user $u$ from using all the services of $R$.

As is generally accepted, maximizing user satisfaction constitutes the key issue to every service. This forces every play $p_i$ to seek to maximize its contribution to the total worth earned by $u$. But this happens under the following restriction. The user $u$ has a limited time to spend interacting with the services and thus he must split this into his needs wisely in order to acomplish his different tasks through $R$. This means that no play can monopolize all the available time of $u$. Moreover, $u$'s multitasking abilities are limited by nature. So spending more time on service $i$ might, on the one hand raise satisfaction from $i$ but on the other might lower satisfaction from service $j$ ($j \neq i$). A mathematical model that describes such interactions among $p_i$'s is the Cournot competition \cite{CournotTheorem}. Under this model it is assumed that the plays do not cooperate with each other but instead decide independently and at the same time about how much value they should produce for $u$. Thus if with $\pi_i$ we declare the profit of each $p_i$ we will have that: 

\begin{align}
\label{ServiceProfit} \pi_i= \max\limits_i (\textit{value generated from $i$-th play} - \textit{cost spent})
\end{align}

Assume that $u$ in play $i$ creates $q_i$, $i=1,\cdots,n$ units of value. In economics usually the function that describes the total value generated from the $i$-th play in its simplest form is given by $(a-\sum\limits_{j=1}^n q_j)q_i$. Here $a$ is a positive constant that represents the size of the environment into which interactions happen. For our model, $a$ could e.g. represent the total size of data held by the resource available for use or the total time that $u$ wants to dedicate to $R$. If, for example, $a$ represents the time available, then the demand function intuitively says that the more value is generated by all the plays the less time remains to be used and vice versa. Since now the cost spent for $i$-th play is $cq_i$ using the above in (\ref{ServiceProfit}) we will have:

\begin{align}
\label{ProfitMaximizer} \pi_i=\max \limits_i (a-\sum\limits_{j=1}^n q_j-c)q_i \text{  }, i=1,\cdots,n
\end{align}

The solutions of (\ref{ProfitMaximizer}) are the Nash equilibria of the plays of $u$ in the case that these plays do not cooperate. These equilibria will help us find the worth of $u$ and reason about his loyalty to $R$. It is easy to prove (the proof can be found in the Appendix, or for a more general case, in \cite{CournotTheorem}) that the solutions of (\ref{ProfitMaximizer}) give the following worth for each play $p_i$ in the case that these act independently (non-cooperatively):

\begin{align}
\label{ProfitOf_i} \pi_i=\left(\frac{a-c}{n+1}\right)^2
\end{align}

Assume now that all the $p_i$'s decide to act in a cooperative manner. In this case due to symmetry we can imagine all the plays combined together into a unique play, so (\ref{ProfitMaximizer}) collapses into a single equation the maximization of which gives:

\begin{align}
\label{GrandCoalition} v(N) \equiv \frac{(a-c)^2}{4}
\end{align}

\noindent where with $v(N)$ we denote the total worth produced when all plays cooperate. Before applying our model to quantify $u$'s loyalty to $R$ let us discuss what cooperation and non-cooperation means for the plays $p_i$. When we say that a set of plays cooperate, we mean that they do not harm each other but instead act as a unity to produce a common value. On the other hand, when there is non-cooperation each play does not care about the other plays but wants only to maximize its own value. Since this idea might sound a bit subtle we give the following example: In the case of Zoogle+ if the 3 services acted in a cooperative manner then their total worth according to (\ref{GrandCoalition}) would be $\frac{(a-c)^2}{4}$. If, on the other hand, they acted in a non-cooperative manner, from (\ref{ProfitOf_i}) each would be worth $\left(\frac{a-c}{3+1}\right)^2$ and their total worth would be $3 \left(\frac{a-c}{4}\right)^2 < \frac{(a-c)^2}{4}$. So it is clear that when under cooperation the plays produce more. 

\section{User Loyalty}

Let us now turn our attention to the loyalty of $u$ to $R$. As said earlier $p_i$'s can act in a cooperative or non-cooperative manner in order to achieve their goals. Apart from total cooperation and non-cooperation there exist intermediate situations in which some of the plays might decide to cooperate and some others will decide to deviate (non-cooperate). It is exactly these cases that will shed some light on $u$'s loyalty to $R$. Consider the following scenario: User $u$ is thinking of abandoning a non-empty set $S \subset N$ of services because he discovered that on a different resource $R'$ he earns $v(S)$ from these. In order to make his final decisions he first reasons as follows: "If I stay loyal to $R$, my total worth from (\ref{GrandCoalition}) is $v(N)$ and on average I earn $\frac{v(N)}{n}$. On the other hand, if I partially switch to $R'$ I would earn on average $\frac{v(S)}{s}$, where $|S|=s$. So I partially switch to $R'$ if $\frac{v(S)}{s} > \frac{v(N)}{n}$." The idea just described describes the notion of the core in the cooperative game played by $p_i$, $i=1,\cdots,n$. We say that the core is non-empty when for any non-empty set of services chosen, $u$ on average earns less than if all the services were chosen. In other words, the core is non-empty when: 

\begin{align}
\label{CoreNonEmpty} &\frac{v(S)}{s} \leq \frac{v(N)}{n}, \forall S \subset N
\end{align}

So finally the user partially deviates from $R$ if he discovers at least one set of services $S^*$ that violates (\ref{CoreNonEmpty}). For this set we will have that $\frac{v(S^*)}{s^*} > \frac{v(N)}{n}  \Rightarrow v(S^*)>\frac{s^*(a-c)^2}{4n}$. In this case $u$ would have an incentive to partially leave $R$.

\subsection{Resource provider's view}

Our approach is not only valuable to users but also to a resource provider. This is because under our model the provider can adopt some strategies to fight a potential rejection of a user. Using equation (\ref{CoreNonEmpty}) we see that the provider must try to keep the core of the game non-empty. So first of all the provider must decide to provide its services in a cooperative way. We can observe this trend in many resource providers nowadays. For example, Google recently (March 2012) unified its services in order to act cooperatively. In this way there is the trend that any two services will cooperate in order to produce a common value. Thus through Gmail you can view or process your attachments through GoogleDocs or use the Dashboard to synchronize all your e-mail contacts with your Android device. In this manner Google tries to keep its core non-empty and thus give incentives to users for more satisfaction. Of course there is always the case of a user using the services in a non-cooperative manner, but then from (\ref{ProfitOf_i}) he would earn less. Moreover, the provider should try to increase the ratio $\frac{(a-c)^2}{4n}$. This can be done in the following ways: First, the provider can help $u$ to reduce his cost $c$. This can be achieved in many ways, e.g. by upgrading its hardware, by hiring a qualified service \cite{AnEvaluationMethodOfOutsourcingServicesForDevelopingAnElasticCloudPlatform}, by adopting process completeness strategies \cite{ProcessCompletenessStrategiesForAligningServiceSystemsWithCustomersServiceNeeds} or by improving the service tutorials and introducing online help desks \cite{AnIntegratedHelpDeskSupportForCustomerServicesOverTheWorldWideWebACaseStudy,WebBasedIntelligentHelpdeskSupportEnvironment}. 
Second, it might consider increasing the factor $a$, which as we said might represent the size of data or the time available. For example, a social network might strive to attract more users thus increasing $a$ so that the current users will belong to a bigger society and become more satisfied, resulting in more options for interaction. The same idea also applies to the critical mass of Service Overlay Networks \cite{CapacityAllocationInServiceOverlayNetworks}. Finally, it might consider reducing the number of services it provides (reducing the denominator of (\ref{CoreNonEmpty}) increases the fraction), for example by obsoleting outdated services or not so popular ones. 

Another important factor for a provider is to collect user rating information data for its services so as to compute its own estimations of how satisfied the users are. According to (\ref{CoreNonEmpty}) the closer the provider's estimation $v_{provider}(S)$ is to the user's one $v_{user}(S)$, the more an effective strategy can be adopted to maintain its customers, and this is because in this way the provider has a clear image of what its users like. Also providers should follow user trends to estimate the potential $S^*$'s that make its resource vulnerable either by asking for feedback from the users or by outsourcing this task to experts \cite{OUTSOURCINGEVALUATIONaProfitableProcess,HowTheTransactionCostAndResourceBasedTheoriesOfTheFirmInformOutsourcingEvaluation}.
Moreover, more complex scenarios can be adopted from the provider's point of view in order to further refine its strategies. For this the provider can design more complicated games into which users would engage themselves. For example, the provider will not only consider how to satisfy the user but also how to earn more money, so instead of having the players compete in order to find the equilibria between value produced and cost spent we could have players competing between satisfaction offered and money earned and cost spent. 

\section{Discussion and future work}

As stated previously, the value generated by the players and the cost spent are related to the user's loyalty to the resource. But can the above be calculated by the user? After all, there are many controversial metrics that can be used for rating. For example, user $u$ in the Zoogle+ example might have various concerns when using the resource: how user friendly are Zoogle+'s interfaces? How much storage space am I allowed? What are the privacy policies of Zoogle+? Do most of my friends belong to Zoogle+ too? And there are even more. An approach to these concerns is the following: On the one hand, the value generated through a user - service interaction must be perceived as a combination A) of the amount of information in the form of data created, exchanged, stored, or retrieved by the user, and B) of the user's personal metrics. For example one such metric is the one we adopted as a cost factor, i.e. the amount of the user's time invested to produce the value through interaction (programming, typing, asking queries, etc.). And this is something natural, but other functions can be used too, such as money spent by the user, bandwidth or CPU resources used, or a combination of the above. One can consult \cite{DynamicSelectionMechanismForQualityOfServiceAwareWebServices} and most of the references therein for a recent treatment of QoS properties and measurement metrics of services.

In our cooperative game we used as a notion of fairness one in which the value generated by the players must split equally among them. This is called a game with transferable utilities since user $u$ makes the decision based on the average value calculated by all the plays. This means that the transfering of profit between any two services $x, y$ is allowed, so as to maintain the same average. But since there exist many different notions of fairness such as the Shapley Value, it is of particular interest to extend the analysis to these notions as well.

Finally, since under our model we assumed that each service is somewhat of the same nature as every other service, in a more realistic scenario in which the services differ, we could have used a differentiation parameter $\gamma$ and the demand from service $i$ would change to $(a-q_i-\gamma\sum\limits_{j=1, j \neq i}^n q_j)$, thus resulting in a more complex worth function. This case is a subject of ongoing work.

\section*{Appendix}

We prove below the Nash equilibria for the game described in section 2.2. The first order derivative conditions of (\ref{ProfitMaximizer}) $\forall i, i=1,\cdots,n$ give:

\begin{align}
\nonumber \frac{\partial }{\partial q_i} [(a-\sum\limits_{j=1}^n q_j-c)q_i]=a-\sum\limits_{j=1, j\neq i}^n q_j-c-2q_i=0 \Rightarrow q_i=\frac{a-c-\sum\limits_{j=1, j \neq i}^n q_j}{2}
\end{align}

The system gives: $\tilde{q} \equiv q_1=\cdots=q_n$, thus $\tilde{q}=\frac{a-c-(n-1)\tilde{q}}{2}$ or $\tilde{q}=\frac{a-c}{n+1}$. Now plugging $\tilde{q}$ in (\ref{ProfitMaximizer}) gives the result.

\end{document}